\documentclass[12pt,preprint]{aastex}

\usepackage{emulateapj5}

\begin{document}
\def\etal{{et al.}}

\slugcomment{Accepted by the Astronomical Journal}

\title{Evidence for Reionization at $z \sim 6$:
Detection of a Gunn-Peterson Trough in a $z=6.28$ Quasar\altaffilmark{1,2}}

\author{Robert H. Becker\altaffilmark{\ref{UCDavis},\ref{IGPP}},
Xiaohui Fan\altaffilmark{\ref{IAS}},
Richard L. White\altaffilmark{\ref{STScI}},
Michael A. Strauss\altaffilmark{\ref{Princeton}},
Vijay K. Narayanan\altaffilmark{\ref{Princeton}},
Robert H. Lupton\altaffilmark{\ref{Princeton}},
James E. Gunn\altaffilmark{\ref{Princeton}},
James Annis\altaffilmark{\ref{Fermilab}},
Neta A. Bahcall\altaffilmark{\ref{Princeton}},
J. Brinkmann\altaffilmark{\ref{ApachePt}},
A. J. Connolly\altaffilmark{\ref{Pittsburgh}},
Istv\'an Csabai\altaffilmark{\ref{JHU},\ref{Eotvos}},
Paul C. Czarapata\altaffilmark{\ref{Fermilab}},
Mamoru Doi\altaffilmark{\ref{UTokyo}},
Timothy M. Heckman\altaffilmark{\ref{JHU}},
G. S. Hennessy\altaffilmark{\ref{USNO}},
\v{Z}eljko Ivezi\'{c}\altaffilmark{\ref{Princeton}},
G. R. Knapp\altaffilmark{\ref{Princeton}},
Don Q. Lamb\altaffilmark{\ref{Chicago}},
Timothy A. McKay\altaffilmark{\ref{Michigan}},
Jeffrey A. Munn\altaffilmark{\ref{Flagstaff}},
Thomas Nash\altaffilmark{\ref{Fermilab}},
Robert Nichol\altaffilmark{\ref{CMU}},
Jeffrey R. Pier\altaffilmark{\ref{Flagstaff}},
Gordon T. Richards\altaffilmark{\ref{PennState}},
Donald P. Schneider\altaffilmark{\ref{PennState}}, 
Chris Stoughton\altaffilmark{\ref{Fermilab}},
Alexander S. Szalay\altaffilmark{\ref{JHU}},
Anirudda R. Thakar\altaffilmark{\ref{JHU}},
D. G. York\altaffilmark{\ref{Chicago},\ref{Fermi}}
}

\altaffiltext{1}{Based on observations obtained at the W.M. Keck Observatory, 
which is operated as a scientific partnership
among the California Institute of Technology, 
the University of California and the National Aeronautics and
Space Administration, made possible by the generous 
financial support of the W.M. Keck
Foundation,  and  with the
Sloan Digital Sky Survey,
which is owned and operated by the Astrophysical Research Consortium.}
\altaffiltext{2}{
This paper is dedicated to the memory of Arthur F. Davidsen (1944--2001), a
pioneer in the study of the intergalactic medium and a leader in the development of the Sloan
Digital Sky Survey.
}
\altaffiltext{3}{Physics Department, University of California, Davis,
CA 95616
\label{UCDavis}}
\altaffiltext{4}{IGPP/Lawrence Livermore National Laboratory, Livermore,
CA 94550
\label{IGPP}}
\altaffiltext{5}{Institute for Advanced Study, Olden Lane,
Princeton, NJ 08540
\label{IAS}}
\altaffiltext{6}{Space Telescope Science Institute, Baltimore, MD 21218
\label{STScI}}
\altaffiltext{7}{Princeton University Observatory, Princeton,
NJ 08544
\label{Princeton}}
\altaffiltext{8}{Fermi National Accelerator Laboratory, P.O. Box 500,
Batavia, IL 60510
\label{Fermilab}}
\altaffiltext{9}{Apache Point Observatory, P. O. Box 49,
Sunspot, NM 88349
\label{ApachePt}}
\altaffiltext{10}{Department of Physics and Astronomy,
          University of Pittsburgh,
          Pittsburgh, PA 15260
\label{Pittsburgh}}
\altaffiltext{11}{
Department of Physics and Astronomy, The Johns Hopkins University,
   3701 San Martin Drive, Baltimore, MD 21218, USA
\label{JHU}}
\altaffiltext{12}{Department of Physics of Complex Systems,
E\"otv\"os University,
   P\'azm\'any P\'eter s\'et\'any 1/A, Budapest, H-1117, Hungary
\label{Eotvos}
}
\altaffiltext{13}{Department of Astronomy and Research Center 
  for the Early Universe, School of Science, University of Tokyo, Hongo,
  Bunkyo, Tokyo, 113-0033, Japan
\label{UTokyo}}
\altaffiltext{14}{U.S. Naval Observatory, 
3450 Massachusetts Ave., NW, 
Washington, DC  20392-5420
\label{USNO}}
\altaffiltext{15}{University of Chicago, Astronomy \& Astrophysics
Center, 5640 S. Ellis Ave., Chicago, IL 60637
\label{Chicago}}
\altaffiltext{16}{University of Michigan, Department of Physics,
        500 East University, Ann Arbor, MI 48109
\label{Michigan}}
\altaffiltext{17}{U.S. Naval Observatory, Flagstaff Station,
P.O. Box 1149,
Flagstaff, AZ  86002-1149
\label{Flagstaff}}
\altaffiltext{18}{Dept. of Physics, Carnegie Mellon University,
     5000 Forbes Ave., Pittsburgh, PA-15232
\label{CMU}}
\altaffiltext{19}{Department of Astronomy and Astrophysics,
The Pennsylvania State University,
University Park, PA 16802
\label{PennState}}
\altaffiltext{20}{Enrico Fermi Institute, 
5640 S. Ellis Ave., Chicago, IL 60637
\label{Fermi}}

\begin{abstract}
We present moderate resolution Keck spectroscopy of 
quasars  at $z=5.82$, 5.99 and 6.28, discovered by the Sloan Digital
Sky Survey (SDSS).
We find that the Ly$\alpha$ absorption in the spectra of these
quasars evolves strongly with redshift. 
To $z \sim 5.7$, the Ly$\alpha$ absorption evolves as expected from an
extrapolation from lower redshifts.  However, in the highest redshift object, 
SDSSp J103027.10+052455.0 ($z=6.28$), 
the average transmitted flux is $0.0038 \pm 0.0026$ times that of
the continuum level
over 8450 \AA\ $<  \lambda <$ 8710\AA\ ($5.95 < z_{abs} < 6.16$),
consistent with zero flux.  Thus the 
flux level drops by a factor of $> 150$, and 
is consistent with zero flux in the Ly$\alpha$
forest region immediately blueward of the Ly$\alpha$ emission line,
compared with a drop by a factor of $\sim 10$ at $z_{abs} \sim 5.3$.
A similar break is seen at Ly$\beta$; because of the decreased
oscillator strength of this transition, this allows us to put a
considerably stronger limit, $\tau_{eff} > 20$, on the optical depth to
Ly$\alpha$ absorption at $z=6$.

This is a clear detection of a complete Gunn-Peterson trough,
caused by neutral hydrogen in the intergalactic medium.
Even a small neutral hydrogen fraction in the intergalactic medium 
would result in an undetectable flux in the Ly$\alpha$
forest region. 
Therefore, the existence of the Gunn-Peterson trough by itself does not 
indicate that the quasar is observed prior to the reionization epoch. 
However, the fast evolution of the mean absorption in these high-redshift 
quasars suggests that the mean ionizing background along the line of sight
to this quasar has declined significantly from $z\sim 5$ to 6,
and the universe is approaching the reionization epoch at $z\sim 6$.

\end{abstract}


\clearpage
\section{Introduction}

Recent discoveries of quasars at redshifts of 5.8 and greater
(\citealt{Fan2000}, \citealt{PaperI}, hereafter Paper I) are
finally allowing quantitative studies of the status of
the intergalactic medium (IGM) and the history of reionization at
redshifts near 6.
The absence of a Gunn-Peterson trough (Shklovsky 1964, Scheuer 1965, 
Gunn \& Peterson 1965) in the spectrum of the $z=5.8$ quasar 
SDSSp J104433.4--012502.2 (SDSS 1044--0125 for 
brevity, \citealt{Fan2000}; see Goodrich \etal\ 2001 for an updated
redshift based on the CIV line in the near-infrared)
indicates that the intergalactic medium (IGM) is already highly
ionized at $z\sim 5.5$, presumably by the UV ionizing photons from
quasars and star-forming galaxies at high redshift.
In Paper I, we used the low resolution discovery spectra of the
three new quasars at $z>5.8$, taken with
the ARC 3.5m telescope at Apache Point Observatory,
to show that Ly$\alpha$ absorption 
increases significantly from a redshift of 5.5 to a redshift of 6.0.
In particular, the spectrum of the $z=6.28$ quasar 
SDSSpJ 103027.10+052455.0 shows that 
in a $\sim 300$\AA\ region of the Ly$\alpha$ forest immediately
blueward of the Ly$\alpha$ emission line, the flux level is consistent
with zero, indicating a flux decrement of $\gtrsim 50$, and suggesting
a possible detection of the complete Gunn-Peterson trough.
To more accurately quantify this effect, 
and to constrain the properties of the IGM,
we have obtained higher resolution spectra of the three new $z>5.8$ quasars 
using the Echelle Spectrograph and Imager (ESI; \citealt{ESI}) on the
Keck II telescope.

In section 2 we describe the observations and present the spectra of
all four $z \gtrsim 5.8$ quasars in the sample of \citet{PaperI}. 
We measure the emission line redshifts of the three new quasars using these
spectra.
In section 3, we discuss the properties of the Ly$\alpha$ forest in the 
four spectra from the point of view of the overall opacity of the IGM
as a function of redshift. 
In section 4
we discuss the cosmological implications of the results.

\section{Spectroscopic Observations}

SDSSp J083643.85+005453.3 ($z=5.82$), SDSSp J130608.26+035626.3 ($z=5.99$)
and SDSSp J103027.10+052455.0 ($z=6.28$,  
referred to as SDSS 0836+0054, SDSS 1306+0356 and
SDSS 1030+0524 throughout this paper for brevity) 
were selected as $i$-dropout objects  
from the Sloan Digital Sky Survey (SDSS, \citealt{York00}) 
multicolor imaging data. 
They were further separated from cool dwarfs using follow-up $J$-band
photometry, and discovery spectra were obtained
using the Double Imaging Spectrograph (DIS), a low-resolution 
spectrograph ($R \sim 500$) on the ARC 3.5m telescope
at the Apache Point Observatory (Paper I). 

In March and May, 2001, 
we obtained moderate resolution spectra of these three high redshift quasars,
using ESI on the Keck II telescope.
The spectra were taken in the echellette mode of ESI. 
In this mode, the spectral range from 4000 to 10000 \AA\ is covered
in ten spectral orders with a constant dispersion of 11.4 km s$^{-1}$ pixel$^{-1}$. 
With a plate scale ranging from 0.142$''$ to 0.128$''$ per pixel in the
three reddest orders, the $1''$ slit has a footprint
of 81--90 km s$^{-1}$ and the $\sim0.7''$ typical seeing has
a footprint of 57--63 km s$^{-1}$.
Using the sky lines in the spectrum, we estimate that the Full Width
at Half Maximum of
the instrument profile is $\sim 1.8$\AA\ at $\sim 8500$\AA, 
indicating a spectral resolution of 66 km s$^{-1}$, or $R\sim 4700$,
almost 10 times higher than that of the discovery spectra.
Wavelength calibration is based on observations of Hg-Ne-Xe lamps 
and Cu-Ar lamps.
The ESI has active flexure control that minimizes any drift 
in the wavelength calibration or fringing pattern. 
This is especially important for accurate sky subtraction --
a crucial step in detecting the faint flux in the quasar Ly$\alpha$
forest region in the presence of strong sky emission lines.
The spectrophotometric standard G191-B2B (\citealt{Massey98},
\citealt{MG90}) was observed for flux calibration.
The standard star, observed at similar airmass as the
quasars, was also used for tracing the spectra in the spectral region where
little flux is detected from the quasar due to strong Ly$\alpha$ absorption.
All the observations were made at the parallactic angle. 
Data reduction made use of standard
IRAF routines along with a suite of custom analysis routines
that handle the large curvature, highly non-linear photometric
response, and slight tilt of the sky lines in the ESI instrument. 
The dates and durations of the exposures are given in Table 1.

\begin{figure*}
\epsscale{0.8}
\plotone{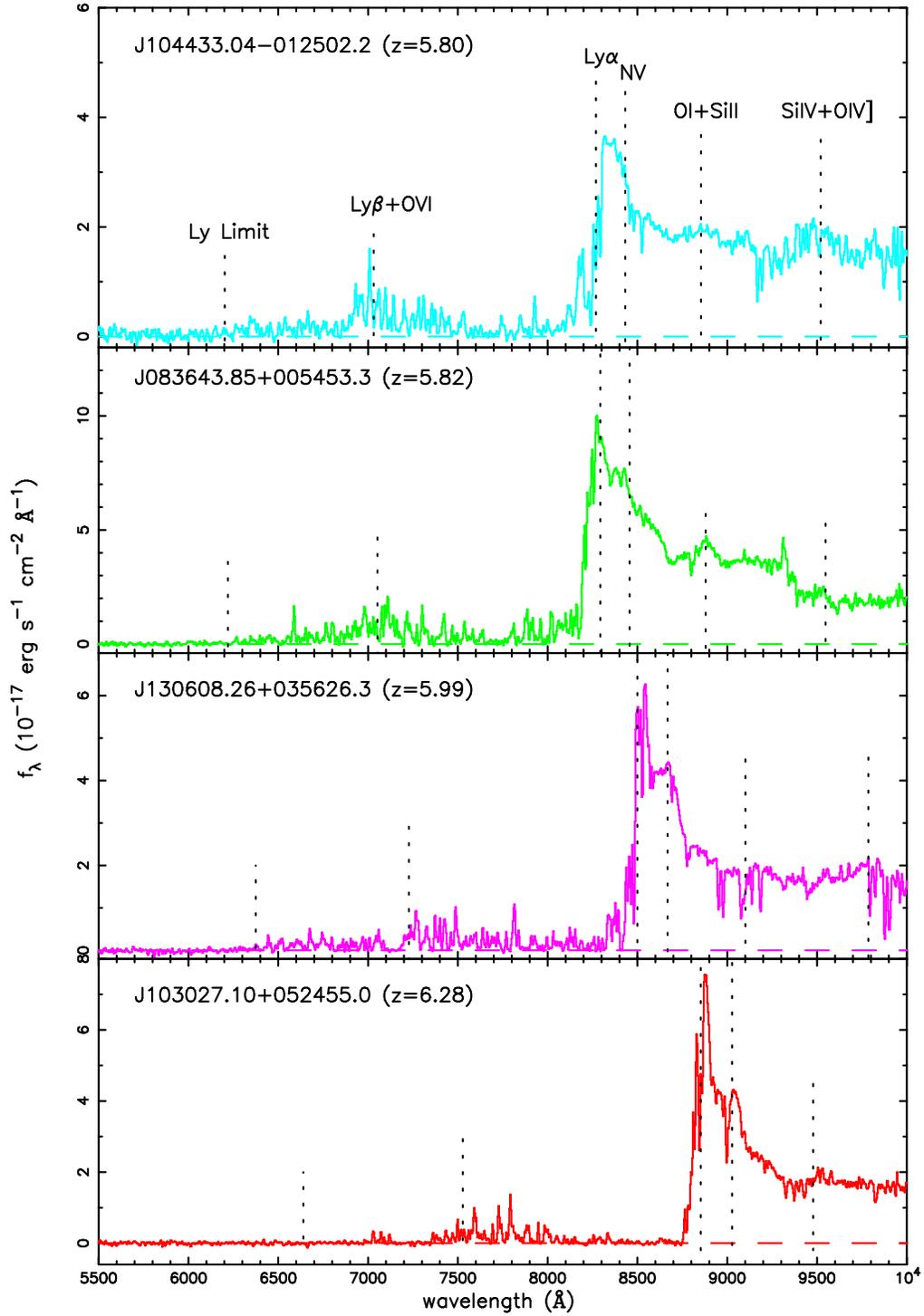}
\caption{
Optical spectra of $z\gtrsim 5.8$ quasars observed with Keck/ESI,
in the observed frame. The spectra have been smoothed to 4\AA\
pixel$^{-1}$, and have been normalized to the observed $z$ band flux.
The spectrum of SDSS1044--0125 has been taken from \citet{Fan2000}.
In each spectrum, the expected wavelengths of prominent emission 
lines, as well as the Lyman limit, are indicated by the dashed lines. 
\label{figspec}
}
\end{figure*}

The three spectra are displayed in Figure~\ref{figspec}, 
along with the spectrum of SDSS 1044--0125,
which was observed an year earlier with the same instrument setting
(Fig.~2 of \citealt{Fan2000}). 
The spectra are binned to 4\AA\ pixel$^{-1}$ to improve the
signal-to-noise ratio, and
the flux levels are  adjusted to match the $z$ band photometry presented in
Paper I.
Each pixel is weighted by the inverse square of the estimated noise on that pixel
when the high resolution spectrum is binned to low resolution.
It is clear from a visual inspection of the four spectra that 
the flux level blueward of Ly$\alpha$ decreases with increasing
redshift, and is consistent with zero flux in SDSS 1030+0524.

Using these spectra, we first determine accurate redshifts from
the emission lines.
As discussed in Fan \etal\ (2000) and in Paper I, 
redshift determination of $z>5.5$ quasars using optical spectra
is difficult, due to the weakness of the accessible metal emission lines 
and the effect of Ly$\alpha$ forest absorption on the Ly$\alpha$
emission line.
The presence of associated metal absorption lines would also
bias the result (\citealt{Goodrich01}).
The strong CIV emission line, now at $\lambda > 10000$\AA, would provide
a more reliable redshift measurement, but it lies beyond our
spectral coverage for all our objects.

{\bf SDSS 1044--0125.}  As mentioned above, the displayed spectrum is
from Fan \etal\ (2000).
\cite{Djorgovski2001} present a higher signal-to-noise
spectrum of SDSS 1044--0125, which shows the increasing optical depth
of the Ly alpha forest with redshift particularly well.

{\bf SDSS 0836+0054.} A strong OI+SiII$\lambda$1302 emission line
is detected at $\lambda = 8873 \pm 1$\AA, with a rest-frame
equivalent width (EW) of 4.2$\pm 0.1$\AA. The Ly$\alpha$+NV emission
line is very broad, with the peak wavelength of the Ly$\alpha$
component at $\sim 8230$\AA, and EW = $70 \pm 3$\AA.
We adopt a redshift of 5.82,  using the central wavelength of OI line,
and a redshift error of 0.02, considering the uncertainty using
a single line for redshift.
The peak of the Ly$\alpha$ emission line is consistent with this redshift.
The EW of OI+SiII is comparable to that of the average of $z\sim 4$ quasars
(3.2\AA, \citealt{SSG91}).

{\bf SDSS 1306+0356.} 
The strong Ly$\alpha$+NV emission line shows a separate NV$\lambda$1240
component. We determine the central wavelengths and
EWs of Ly$\alpha$ and NV by fitting two Gaussian profiles to the 
wavelength range redward of the peak of Ly$\alpha$ emission, i.e.,
the region not affected by the Ly$\alpha$ forest absorption.
The wavelength ratio of the two components is fixed in the fitting procedure.
We find $z_{em} = 5.99$, EW(Ly$\alpha$) = 38.1 $\pm$  14.9\AA, 
and EW(NV) = 17.9 $\pm$ 6.3 \AA. 
No obvious OI+SiII emission line is detected; any emission line there
is probably affected by the CIV absorption feature at the same wavelength 
(see below).
A possible SiIV$\lambda$1402 feature is detected at $\sim 9800$\AA, but
it is difficult to fit its profile due to the weakness of the line and
possible absorption lines nearby.
We therefore adopt a redshift of $5.99 \pm 0.02$ for SDSS 1306+0356.

In the spectrum of SDSS 1306+0356, we notice a strong absorption
feature at $\sim 7130$\AA, where over $\sim 80$\AA, there is no detectable
flux. The rest-frame equivalent width is $\sim 15$\AA, typical for a 
damped Ly$\alpha$ system, at a redshift of $z_{abs} = 4.86$.
A strong absorption feature is detected at $\lambda = 9080$\AA, 
corresponding to CIV absorption at the same redshift.  
This feature is double peaked in absorption, consistent with the 
$\lambda\lambda$1548, 1551 components of the CIV doublet, although
the signal-to-noise ratio is low at that wavelength.
This system, if confirmed by high signal-to-noise ratio spectroscopy, 
is the highest-redshift
damped Ly$\alpha$ system known (the previous record holder was at
$z=4.47$; P\'eroux \etal\ 2001; Dessauges-Zavadsky \etal\ 2001).
The two other doublet absorption features at $\sim 9900$\AA\ and
$\sim 8960$\AA, are identified as MgII absorptions at $z=2.53$ and 2.20,
respectively.
Note that other spectra in this paper also show various absorption features
redward of Ly$\alpha$ emission. The detailed identifications of these
metal lines are beyond the scope of this paper.

{\bf SDSS 1030+0524.} 
In Paper I, we presented the Keck/NIRSPEC $J$-band spectrum of this object,
in which a strong CIV feature (EW = $31.5 \pm 8.6$\AA) is detected,
at $z=6.28 \pm 0.02$.
The optical spectrum shows a separate NV emission line, and possible
detection of OI+SiII and SiIV lines at the same redshift.
We measure EW(Ly$\alpha$) = 40.9 $\pm$ 7.4\AA, and EW(NV) = 16.9 $\pm$ 4.0 \AA,
by fitting Gaussian profiles to the blended Ly$\alpha$+NV emission line.

Paper I discusses the implication of the detection of metal emission lines,
in particular, that of NV emission, to the metallicity of the quasar
environment. Assuming a power law continuum of $f_\nu \propto \nu^{-0.5}$,
the flux ratio NV/CIV is $0.74 \pm 0.27$, consistent with a super-solar
metallicity $Z \gtrsim 3 Z_{\odot}$, according to the calculations
of Hamann \& Ferland (1993).
Note, however, these calculations assume a single-zone photo-ionization model,
and might be affected by the uncertainties in heavy element ratios, as well
as possible enhancement of NV flux  by resonant
scattering from the red wing of the Ly$\alpha$ emission line
(\citealt{KV98}).

\section{Neutral Hydrogen Absorption}

\begin{figure*}
\epsscale{0.8}
\plotone{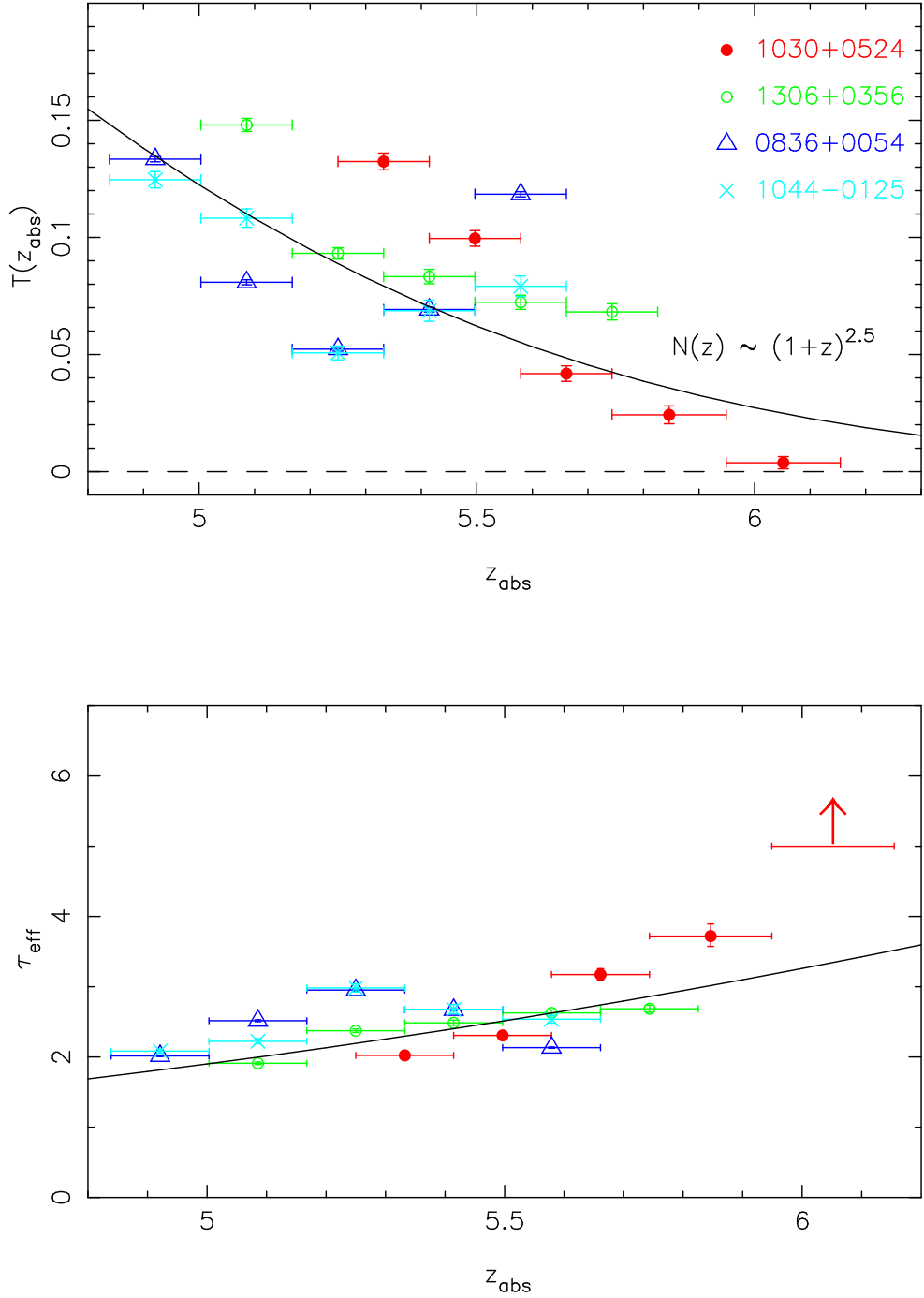}
\caption{
Evolution of transmitted flux ratio and effective Gunn-Peterson
optical depth as functions of redshift.
The solid line is the expected evolution if the number density of
Ly$\alpha$ clouds increases as $N(z) \propto (1+z)^{2.5}$.
No flux is detected in the spectrum of SDSS 1030+0524 at $z_{abs} \sim 6$,
indicating $\tau_{eff} > 5.0$.
The non-detection of flux in the Ly$\beta$ trough gives a
substantially stronger $1\sigma$ upper limit of $\tau_{eff} > 20$.
\label{figtrans}
}
\end{figure*}

In Paper I, we calculated the average absorption in the
Ly$\alpha$ forest region based on the ARC 3.5m discovery spectra.
Here we carry out these calculations with the Keck spectra.
The results are summarized in Table 2.
The table includes the measurement of the $D_{A}$ and $D_{B}$
parameters following Oke \& Korycansky (1982),
the average flux decrement in the Ly$\alpha$ and Ly$\beta$ forest
regions, respectively, as well as the transmitted flux ratio 
$[{\mathcal T}(z_{abs})]$ defined in Paper I, 
measured in the highest redshift window in each quasar 
that is not affected by the Ly$\alpha$ emission line and the proximity effect.
The definitions of these quantities are given in Paper I.
The transmitted flux ratio is measured in the redshift range 
$z_{em} - 0.4 < z_{abs} < z_{em} - 0.2$ for all objects except
SDSS 1030+0524, for which the range is $5.95 < z_{abs} < 6.16$.
Additional measurements of this quantity along the line of
sight to each quasar at lower redshift are illustrated in Figure~\ref{figtrans}. 
All the measurements here and below assume an intrinsic quasar continuum
$f_\nu \propto \nu^{-0.5}$ ($f_\lambda \propto \lambda^{-1.5}$), with the
continuum normalized at
rest frame wavelength of $1270 \pm 10$\AA, a region free of major
emission lines.
The unknown continuum shape is a source of systematic error.  However,
for a wavelength window close to the 
Ly$\alpha$ emission, and in the regime in which ${\mathcal T}(z_{abs})
\ll 1$, this error is much smaller than the cosmic variance of
${\mathcal T}$ (see below).  
As in Paper I, the error bars on the $D_{A}$ and $D_{B}$ measurements
reflect the uncertainties of the continuum shape in the
Ly$\alpha$ forest region, while the error bars on ${\mathcal T}(z_{abs})$ 
include only the photon noise. 
In calculating the average flux and the photon noise, 
the signal in each pixel is weighted by the inverse square of the
estimated noise level in that pixel.
The values in Table 2 are consistent with the measurements in Paper I,
albeit with errors lower by a factor of $\sim 5$.

In Table 2, we also include the measurement of the redshift of
the Lyman Limit System ($z_{LLS}$), following the method described
in Fan \etal\ (2000, \S4.1). Note, however, at $z\sim 6$, the
flux level at the rest-frame 912\AA\ is so low due to the
presence of numerous Ly$\alpha$ forest lines from lower redshift, that this
measurement is rather uncertain.
The quasar spectra are consistent with no flux almost immediately
blueward of rest-frame 912\AA.

Figure~\ref{figtrans} presents the evolution of the transmitted flux ratio 
${\mathcal T}$ and effective optical depth 
$\tau_{eff} (\equiv -\ln({\mathcal T}))$
as a function of redshift $z_{abs}$.
The average absorption evolves strongly with redshift.
At $z_{abs} \sim 3.5$, ${\mathcal T} \sim 0.4$ (e.g. \citealt{Rauch97}).
It decreases to ${\mathcal T} \sim 0.25$ at $z\sim 4.5$ (\citealt{Songaila99}).
At $z\sim 5.0$, $\langle {\mathcal T} \rangle \sim 0.12$ from Figure~\ref{figtrans},
and at $z \sim 5.5$, $\langle {\mathcal T} \rangle \sim 0.07$.
The scatter of ${\mathcal T}$ at the same redshift is larger than implied by
photon noise or uncertainty of the continuum shape, suggesting that
the error in  ${\mathcal T}$ is dominated by cosmic variance. 
Zuo (1993) calculated the scatter of the transmitted flux
given the number density and column density distribution of Ly$\alpha$
clouds. 
Using Eq. (9) in that paper, we find that at $z\sim 5.5$, 
for ${\mathcal T} \sim 0.10$,
$\sigma({\mathcal T}) \sim 0.03$, comparable to  the scatter seen in
Figure~\ref{figtrans}.
For ${\mathcal T} \ll 1$, $\sigma({\mathcal T})$ is comparable to 
${\mathcal T}$ itself, and decreases rapidly toward higher redshift. 

The only measurements of ${\mathcal T}$ at $z>5.7$ come from the highest 
redshift object SDSS 1030+0524.
It shows that ${\mathcal T}(z_{abs} = 5.75) = 0.03$.
The most dramatic change in ${\mathcal T}$, however, comes from the 
measurement at $z\sim 6$; over the redshift range $5.95 < z_{abs} < 6.16$,
the average transmitted flux ${\mathcal T} = 0.0038 \pm 0.0026$.  This
$1.5\,\sigma$ ``detection'' is in fact consistent with zero in our
estimation; indeed, slight changes in the way we did our sky
subtraction caused the value to change by of order $1\,\sigma$.
Thus there is no detectable flux over the wavelength range 
of $8450 < \lambda < 8710$\AA, corresponding to a flux decrement 
of $>150$ from the continuum level, or equivalently, to 
a lower limit (1-$\sigma$) on the Gunn-Peterson optical depth due to 
neutral hydrogen in the IGM of $\tau_{GP} > 5.0$. 
At $\lambda > 8750$\AA,  the flux is detectable again, 
presumably due to the effect of the ionizing
photons from the luminous quasar itself (\S4).
This result confirms the measurement based on the discovery spectrum
(Paper I), in which we found ${\mathcal T} = 0.003 \pm 0.020$.
This is the first detection of a complete Gunn-Peterson trough,
in the sense that no flux is detected over a large wavelength range
in the Ly$\alpha$ forest region, indicating that the effective 
Gunn-Peterson optical depth caused by neutral hydrogen in the IGM,
$\tau_{eff}$, is much larger than one. 

\begin{figure*}
\epsscale{1.0}
\plotone{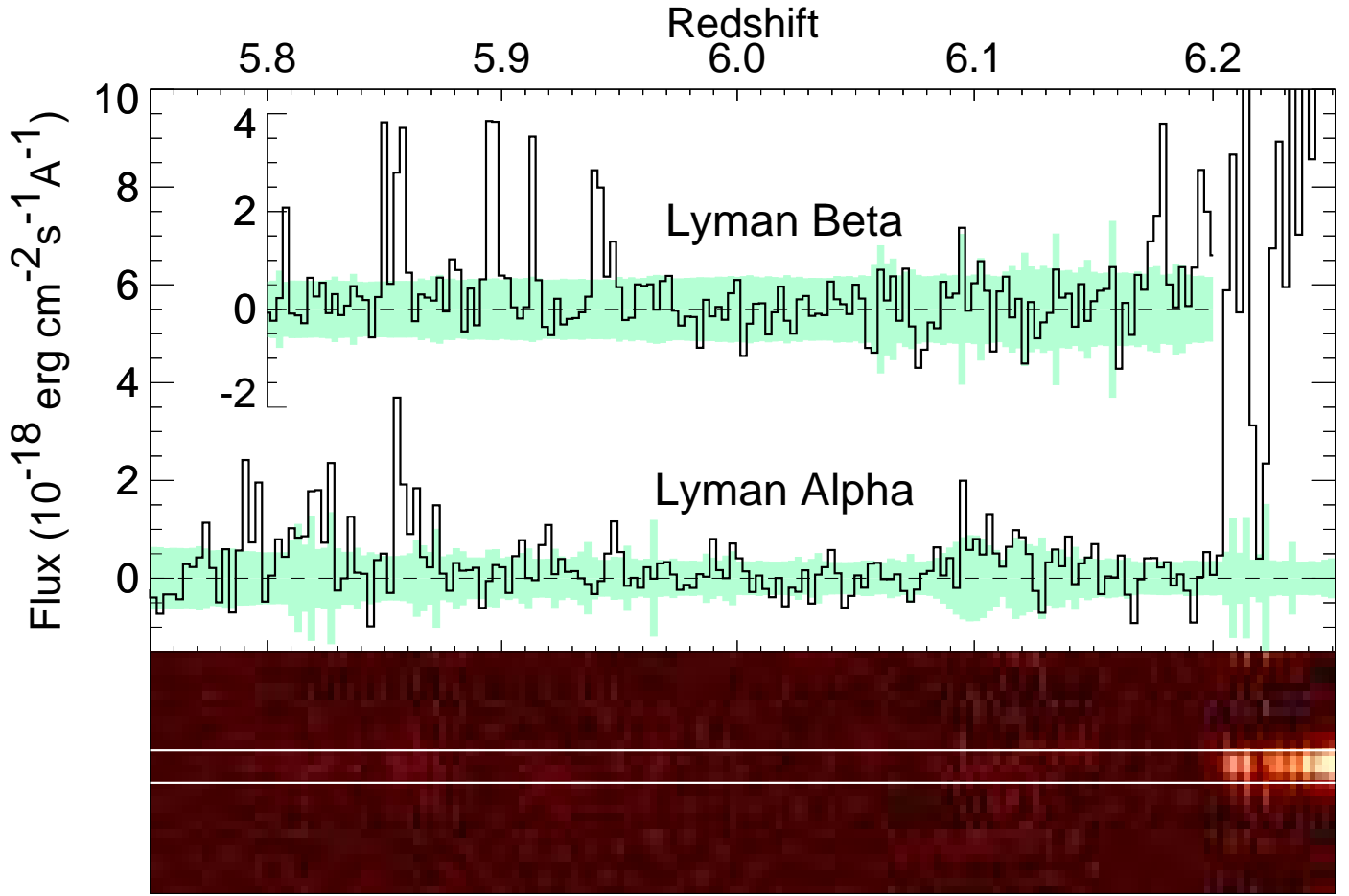}
\caption{
Gunn-Peterson trough in the spectrum of SDSS 1030+0524 observed
with Keck/ESI.
Both the Ly$\alpha$ and Ly$\beta$ absorption regions are shown.
The bottom panel shows the sky-subtracted two-dimensional spectrum in the
Ly$\alpha$ region.  The Ly$\alpha$ data come from the ninth
order of the echellette spectrum and cover the wavelength range
8200\AA\ to 8810\AA.  The Ly$\beta$ data are the result of combining
orders 7 and 8 and cover the wavelength range 6920\AA\ to 7440\AA.
The green shaded areas are the $1\sigma$ noise bands in the extracted
spectra.  The spectrum is very black over the redshift range
$5.95 < z < 6.16$.
\label{fig2d}
}
\end{figure*}

Figure~\ref{fig2d} shows the sky-subtracted two-dimensional spectrogram from the
ninth order of the ESI data; note the complete absence of
detected flux blueward of the Ly$\alpha$ emission line. 
Figure~\ref{fig2d} also shows the one-dimensional spectrum and the corresponding 
estimated error per pixel (shaded) of SDSS 1030+0524 over the
redshift range $5.75<z<6.25$ for both Ly$\alpha$ and Ly$\beta$.
As evident from the figure, the average flux level is consistent
with zero for $5.95 < z < 6.16$ at both lines (8450 -- 8710 \AA\ for
Ly$\alpha$, and 7130 -- 7350 \AA\ for Ly$\beta$).

The effective continuum at Ly$\beta$ is affected by the Ly$\alpha$
forest.  The effective Ly$\alpha$ redshift at 7240 \AA, the center of
the Ly$\beta$ trough, is $z \approx 5$.  Figure~\ref{figtrans} shows
that at that redshift, the transmission of the Ly$\alpha$ forest is
roughly 12\%, albeit with large scatter (see also Songaila \etal\
1999).  Taking this into account, and again extrapolating the quasar
continuum from redward of Ly$\alpha$ emission assuming $f_\lambda
\propto \lambda^{-1.5}$ gives a measured transmitted flux ratio in the
Ly$\beta$ trough of $-0.002 \pm 0.020$, a clear non-detection. 
The error does not include the uncertainty in the transmission of
Ly$\alpha$ at $z=5$, which we estimate to be approximately 20\%
in $\tau_{eff}$ based on the scatter seen in Fig.~\ref{figtrans}.

The ratio of oscillator strengths of the Ly$\alpha$ and Ly$\beta$
lines is 5.27 (Verner, Verner, \& Ferland 1996).  Thus we can convert
the transmitted flux ratio in the Ly$\beta$ forest region 
to an {\em equivalent} $1\sigma$ lower limit to the optical depth
in the Ly$\alpha$ forest region of $\tau_{eff}>20$
(albeit with an uncertainty due to the limits of
our model for the intrinsic spectrum).  This limit is, in fact, 
considerably stronger than the limit obtained directly from the
Ly$\alpha$ forest itself.   

At $\lambda < 6980$\AA, there appears to be another break
(Figure~\ref{figspec}); this is the start of
the corresponding Ly$\gamma$ Gunn-Peterson trough.  We have not
attempted to obtain a quantitative optical depth for this line, due to
the extreme overlapping absorption from both the Ly$\alpha$ and
Ly$\beta$ forests.  

Our ability to constrain the upper limit of the flux in the Ly$\alpha$
and Ly$\beta$ 
forest regions, 
and the corresponding lower limit to the Gunn-Peterson optical depth, 
depends on the reliability of sky subtraction.
In this case, the expression ``black as night'' has special significance 
because invariably, upper limits to brightness come down to the accuracy 
with which one can subtract the  sky. 
Unfortunately, the night sky is not very black, especially in the near IR
region, where the sky emission is concentrated in a series of very bright 
OH lines that nearly blanket the spectrum.
At the ESI resolution, the peak of the strongest sky lines at 
8000 -- 9000\AA\ can
be a factor of $>50$ stronger than that of the darker region between lines.
The high spectral resolution of ESI enables us to resolve the much 
darker regions between strong sky lines and  
limits the number of pixels affected by the sky lines.
The pixels affected by strong sky lines are assigned much smaller weights 
when we calculate the average flux and effective optical depth.
It is our considered opinion that in this instance, the data reduction
posed no unusual difficulties and that the faint to nonexistent flux 
levels seen blueward of Ly$\alpha$ are reliable.  These problems are
less severe in the spectral region of the Ly$\beta$ trough, and as
we've seen, the two troughs give consistent limits
on the mean optical depth.

\section{Discussion}
The Gunn-Peterson (1965) optical depth in a uniformly distributed IGM
is given by
\begin{equation}
\tau_{GP} (z) = 1.8 \times 10^5 h^{-1} \Omega_M^{-1/2}
\left( \frac{\Omega_b h^2}{0.02} \right)
\left ( \frac{1+z}{7} \right )^{3/2}
\left( \frac{n_{\rm HI}}{n_{\rm H}} \right ).
\end{equation}
Only a small fraction of neutral hydrogen component in the IGM is
needed to have $\tau_{GP} \gg 1$ and  give rise to a complete Gunn-Peterson
trough.
Therefore, the existence of a Gunn-Peterson trough by itself does not 
prove that the object is observed prior to the reionization epoch.
Even if the neutral fraction were independent of redshift, one still
expects the transparency of the Ly$\alpha$ forest to decrease with 
increasing  redshift. 
In reality, the expansion of the universe and the increase in the ionizing
background cause
the neutral hydrogen fraction to decrease towards low redshift (see below).
The observed redshift evolution of the Gunn-Peterson optical depth 
in the spectrum of SDSS 1030+0524 allows us to answer an important question:
Is the amount of Ly$\alpha$ forest absorption at $z\sim 6$  consistent
with a simple extrapolation from the observations at lower redshifts? 
Or does it indicate a more dramatic change in the state of the IGM?

Assuming that the number density of Ly$\alpha$ forest lines evolves with
redshift as $(1+z)^\gamma$, where $\gamma \sim 2.5$ at $z\sim 4$,
the effective optical depth evolves as $(1+z)^{\gamma+1}$
(\citealt{ZP93}, \citealt{Z93}).
The solid line in Figure~\ref{figtrans} shows an extrapolation of the optical depths 
to higher redshifts using this simple model. It is evident that 
the $z\sim 6$ value from 
SDSS 1030+0524 deviates from this extrapolation; the observed absorption 
at $z\sim 6$ is stronger than that predicted by this model assuming
an ionization fraction that remains constant with redshift.
A natural explanation for this difference is that  the ionization
fraction increases with time, i.e., the IGM was more  neutral at 
earlier times.

Assuming that the hydrogen gas in the IGM responsible for the Ly$\alpha$ 
forest absorption is in photoionization equilibrium, wherein photoionization
of the neutral hydrogen by the UV background is balanced by recombination,
one can express $\left(n_{\rm HI}/n_{\rm H} \right )$ in 
equation (1) in terms of the ionizing background.
Doing so, the Gunn-Peterson optical depth of a uniform IGM can be 
expressed as (\citealt{Weinberg97}),
\begin{equation}
\tau_{GP} = 5.9 \times 10^{-4} H_0 (1+z)^6 H(z)^{-1} h^{-1} T_4^{-0.7} (\Omega_b h^2/0.02)^2 \Gamma_{-12}^{-1}.
\end{equation}
Here, $H(z)$ is the Hubble constant at redshift $z$, 
$T_4$ is the temperature of the IGM gas in units of 10$^4$K, 
and $\Gamma_{-12}$, the photo-ionization rate of hydrogen in units
of $10^{-12}$ s$^{-1}$, depends on the shape and amplitude of the
ionizing background.
Thus, in a given background cosmology, the observed evolution of the 
Gunn-Peterson optical depth can be used to measure the redshift evolution 
of the ionizing background (e.g., \citealt{McDonald00}).
A more accurate determination of the ionizing background should also
model the gravitational evolution of the Ly$\alpha$ forest with redshift.
Recently, \citet{MM01} used the transmitted flux measurement to calculate the
evolution of ionizing background at $z<5.2$.
In a separate paper (\citealt{Fan2001b}), we 
use both semi-analytic models and cosmological simulations
to investigate the evolution of the ionizing background at higher redshifts,
using the measured transmitted flux ratios and effective optical depths at 
$z \sim 6$.

As discussed by \citet{M1998}, the spectra of sources observed prior
to complete reionization should show the red damped wing of the Gunn-Peterson
trough in the red side of Ly$\alpha$, due to the very large optical depth.
This damped wing would, in principle, suppress the Ly$\alpha$ emission line.
However, as shown in \citet{MR00} and \citet{CH00}, luminous quasars
such as those discussed in this paper would ionize the surrounding regions
and create HII regions of radius several Mpc (the ``proximity'' effect;
\citealt{carswell82}).
The presence of these quasar HII regions results in a much higher
transmission redward of Ly$\alpha$; thus the emission line profile is less 
affected by the red damped wing (see also \citealt{Fan2001b}).
Therefore, the presence of the Ly$\alpha$ emission line, and the absence
of the red damped wing, are not by themselves indicators that the 
IGM was already ionized at this redshift.  The true extent of the proximity
effect in SDSS 1030+0524 is probably best defined by the transmission of
Ly$\beta$ emission at a redshift of 6.18 (see Fig.~\ref{fig2d}).

The size of the HII region associated with the quasar depends on the
luminosity and lifetime of the quasar and the clumpiness of the IGM.
In principle, we could use the proximity effect to constrain the
amplitude of the ionizing background (\citealt{bajtlik88}), by
determining the distance from the quasar at which the optical depth of
the Ly$\alpha$ forest is half that in the Gunn-Peterson trough.
However, we only have a lower limit on the latter, and so at best, we
can only put a lower limit on the distance, and thus an upper limit on
the ionizing background.  We have calculated this upper limit, and
find that it gives a much weaker constraint than 
comparing the data with detailed simulations (Fan et al.~2001b). 
Further, clumpiness of the gas distribution near the quasar on
these length scales will also increase the absorption blueward of
Ly$\alpha$, and hence lead to an artifically higher estimate of the
ionizing background.

Could any local effect produce the apparent Gunn-Peterson trough in the 
spectrum of SDSS 1030+0524? 
A 300\AA\ region at $z\sim 6$ corresponds to a comoving distance of
$\sim 70 h^{-1}$ Mpc ($\Omega = 0.3$, $\Lambda = 0.7$), much larger 
than any large scale structure at that redshift. It is also unlikely 
to be a very strong damped Ly$\alpha$ absorption system, as we do not 
see any corresponding metal lines 
(such as SiIV$\lambda$1402).
We are not aware of any known damped Ly$\alpha$ absorption line at
lower redshifts with an EW of $\sim 300$\AA.
At $z_{abs} \sim 6.05$, it would have a rest frame EW of $\sim 42$\AA,
and $N_{HI} \sim 10^{21.5}$ cm$^{-2}$.
Using the statistics of \citet{APM}, we expect to find only 
0.02 such systems per unit redshift at $z\sim 6$.  Moreover, one would
expect to see associated metal lines from such a strong absorber,
which are clearly not present. 

However, we emphasize that the complete Gunn-Peterson trough
is observed in only one (the highest-redshift) object, and if the
reionization is non-uniform, one may not see exactly the same
dependence of the Gunn-Peterson trough with redshift along other
lines of sight.
In order to gain a more complete understanding of the IGM at $z\gtrsim 6$,
more lines of sight are clearly needed.  The SDSS is expected to find
$\sim 20$ quasars at $z\gtrsim 6$ over the course of the survey (Paper
I), so we can look forward to opportunities to study this question in
detail over the next few years.


\acknowledgements

The Sloan Digital Sky Survey (SDSS) is a joint project
of The University of Chicago, Fermilab, the Institute for
Advanced Study, the Japan Participation Group,
The Johns Hopkins University, the Max-Planck-Institute for
Astronomy (MPIA), the Max-Planck-Institute for Astrophysics (MPA),
New Mexico State University,
Princeton University, the United States Naval Observatory,
and the University of Washington. Apache Point
Observatory, site of the SDSS telescopes,
is operated by the Astrophysical Research Consortium (ARC).
Funding for the project has been provided by the
Alfred P. Sloan Foundation, the SDSS member institutions,
the National Aeronautics and Space Administration,
the National Science Foundation, the U.S. Department of
Energy, the Japanese Monbukagakusho,
and the Max Planck Society. The SDSS Web site is
http://www.sdss.org.
We acknowledge support from 
NSF grants AST-98-02791 and AST-98-02732 and the Institute
of Geophysics and Planetary Physics (operated under the auspices of the
U.S. Department of Energy by Lawrence Livermore National Laboratory
under contract No.~W-7405-Eng-48) (RHB), from
NSF grant PHY00-70928 and
a  Frank and Peggy Taplin Fellowship (XF),
from the Space Telescope Science Institute (RLW),
and from NSF grant AST-0071091 (MAS).
We wish to thank an anonymous referee for correcting an error in the
interpretation of the Ly$\beta$ absorption.
Thanks to David Helfand for his support during the Keck observing
run and for cooking RHB and RLW a fine dinner while we were observing.

\newpage
\begin{deluxetable}{cccc}
\tablenum{1}
\tablecolumns{4}
\tablecaption{Observing Log}
\tablehead
{
Object & redshift($z_{em})$ & Date & Exposure Time (seconds)
}
\startdata
J104433.04--012502.2 & 5.80 & 2000 Mar 04 & 2 $\times$ 1200 \\
J083643.85+005453.3 & 5.82 & 2001 Mar 19 & 1200 \\
J130608.26+035626.3 & 5.99 & 2001 May 26 & 900 \\
J103027.10+052455.0 & 6.28 & 2001 May 26 & 2 $\times$ 900 
\enddata
\end{deluxetable}

\begin{deluxetable}{cccccccc}
\tablenum{2}
\tablecolumns{7}
\tablecaption{Absorption Properties of $z>5.8$ Quasars}
\tablehead
{
object & $z_{em}$ & $z_{LLS}$ & $D_{A}$ & $D_B$ & $z_{abs}$ & Transmitted flux
ratio \\
& & & & &  & at $z_{abs}$
}
\startdata
J104433.04--012502.2 & 5.80 & 5.72 & 0.90 $\pm$ 0.02 & 0.95 $\pm$ 0.02 &  5.5 & 0.088 $\pm$ 0.004 \\
J083643.85+005453.3 & 5.82 & 5.80 & 0.92 $\pm$ 0.02 & 0.95 $\pm$ 0.02 & 5.5 & 0.106 $\pm$ 0.001 \\
J130608.26+035626.3 & 5.99 & 5.98 & 0.91 $\pm$ 0.02 & 0.95 $\pm$ 0.02 & 5.7 & 0.070 $\pm$ 0.003 \\
J103027.10+052455.0 & 6.28 & 6.28 & 0.93 $\pm$ 0.02 & 0.99 $\pm$ 0.01 & 6.05 & 0.004 $\pm$ 0.003\tablenotemark{a}
\enddata
\tablenotetext{a}{The Ly$\beta$ trough gives a $1\sigma$ lower limit to the
equivalent Ly$\alpha$ optical depth of $\tau_{eff}>20$, which corresponds
to an equivalent ${\cal T} < 2\times10^{-9}$.}
\end{deluxetable}

\end{document}